\documentstyle[preprint,aps,tighten]{revtex}
\begin{document}
\draft

\tolerance=5000

\newcommand\be{\begin{equation}}
\newcommand\ee{\end{equation}}
\newcommand\bea{\begin{eqnarray}}
\newcommand\eea{\end{eqnarray}}
\newcommand\nn{\nonumber \\}
\newcommand\e{{\rm e}}

\newcommand\SEH{S_{\rm EH}}
\newcommand\SGH{S_{\rm GH}}

\title{Graviton correlator and metric perturbations in de Sitter 
brane-world}

\author{Shin'ichi Nojiri\thanks{Electronic address: 
snojiri@yukawa.kyoto-u.ac.jp, nojiri@cc.nda.ac.jp}}
\address{Department of Applied Physics,
National Defence Academy,
Hashirimizu Yokosuka 239-8686, JAPAN}

\author{Sergei D. Odintsov\thanks{On leave from Tomsk State 
Pedagogical University, 
634041 Tomsk, RUSSIA.
Electronic address: odintsov@ifug5.ugto.mx}}
\address{
Instituto de Fisica de la Universidad de Guanajuato,
Lomas del Bosque 103, Apdo.\ Postal E-143, 
37150 Leon, Gto., MEXICO}

\author{Sachiko Ogushi\thanks{JSPS fellow, 
Electronic address: ogushi@yukawa.kyoto-u.ac.jp}}
\address{Yukawa Institute for Theoretical Physics, 
Kyoto University, Kyoto 606-8502, JAPAN}

\maketitle
\begin{abstract}

We consider de Sitter brane-world motivated by dS/CFT correspondence
where both bulk and brane are de Sitter spaces. The brane tension is 
fixed by holographic RG. The 4d effective action for metric perturbations 
and 4d graviton correlator are explicitly found. The induced values of
cosmological and Newton constants are calculated. The short distance 
behaviour of the graviton correlator (when no brane matter presents) turns 
out to be significally 
stronger than in the case of General Relativity. It is shown that quantum
brane CFT gives the dominant contribution to graviton correlator on
small scales like in Brane New World scenario.

\end{abstract}


\section{Introduction}

The brane-world idea created the number of new and actively 
developing directions in cosmology. Especially, early time as well
as quantum cosmology should be re-analyzed having in mind the possibility
of large extra dimensions. Clearly, that is fundamental program which is 
still at the very beginning. In such situation (when already the big 
number of brane-world scenarios exists (see \cite{louko} for list of 
references))
only some particular results in frames of specific models may 
be considered.

Among the existing inflationary  brane-world scenarios there is one quantum 
brane-world 
(or Brane New World) \cite{NOZ,HHR} scenario where quantum effects play the 
dominant role. This (AdS bulk) model is formulated in terms of AdS/CFT 
correspondence,
so classical brane tension is fixed from the very beginning.
 CFT living on the brane provides the 
effective tension (via the account of corresponding conformal anomaly), 
giving the possibility of four-dimensional brane inflation. The careful 
calculation of graviton correlator and perturbations in Brane New World 
leads to quite remarkable result\cite{HHR}: CFT strongly suppresses metric 
perturbations at small scale (microwave sky). It is expected that this is 
quite universal phenomenon in quantum cosmology.

It is extremely interesting to check the universality of this phenomenon 
in other brane-worlds. In particular, with recently proposed dS/CFT 
correspondence (see ref.\cite{strominger} for introduction) there 
may be 
necessary to study brane-worlds with de Sitter bulk. One such model 
formulated in the analogy with Brane New World was suggested in 
ref.\cite{SN}. Even in situation when surface counterterms (classical 
brane tension) are fixed
(by requirement of finiteness of de Sitter bulk or by holographic RG 
considerations) one finds the existence of de Sitter brane without
the introduction of effective tension by brane quantum fields \cite{SN}.
Unlike to Brane New World model with AdS bulk, an arbitrary CFT may
be considered on the brane,
the classical dS brane solution will be just modified by quantum effects.  

It is the purpose of the present work to investigate the metric perturbations
and graviton correlator in 5d de Sitter brane-world scenario \cite{SN} using
methods developed in ref.\cite{HHR}. In the next section the formulation of
 dS bulk model with dS brane on it is presented.
The effective action for metric perturbations as well as 4d graviton 
correlator are found. The induced values of 4d cosmological and 
gravitational constants in terms of 5d quantities are calculated.
It is shown that short distance behaviour of the correlator is similar to 
that from 5d AdS bulk (with 4d inflationary brane). In other words,
the induced gravity force is getting stronger than the one in General 
Relativity at short distances. It is observed that when brane tension is 
free parameter it is easiear to realize the inflationary brane in dS bulk.
The graviton correlator is found also in this case when bulk has
 Lorentzian signature. The possibility to have brane radius bigger than 
length parameter of dS bulk is observed.

Section three is devoted to account of  brane CFT effects to graviton
correlator. Using conformal anomaly the CFT action for metric 
perturbations is found. It is shown that CFT gives the dominant contribution
to graviton correlator at short disctances, similarly to the case with 
AdS bulk. Some outlook is given in the Discussion.

\section{Graviton correlator  in de Sitter 
brane-world}

We start from the action $S$ which is the sum of
the Einstein-Hilbert action $S_{\rm EH}$ with
positive cosmological constant, the Gibbons-Hawking
surface term $S_{\rm GH}$ and the first counter term $S_{1}$ :    
\bea
\label{action}
S &=& S_{\rm EH}+S_{\rm GH}+S_{1} ,\quad 
S_{\rm EH} = -{1\over 16\pi G} \int d^{5}x \sqrt{ g }  
\left( R -{12\over l^2}\right)\; , \nn
S_{\rm GH} &=& -{1\over 8\pi G} \int d^{4}x \sqrt{\gamma} 
\; K \; , \quad S_{1}=
{3\over 8\pi G l}\int d^{4}x\sqrt{\gamma} \; .
\eea
Here the 5-dimensional bulk spacetime metric is
given by the $g_{\mu\nu}$ and  the boundary
4-dimensional spacetime  metric is given by $\gamma_{ij}$.
  $K$ is the trace
of the extrinsic curvature of the boundary, which is
defined as\footnote{We adopt the conventions from ref.\cite{HHR}}
\bea
K_{\mu\nu}=\xi^{\rho}_{\mu}\xi^{\sigma}_{\nu}\nabla_{\rho}n_{\sigma}\; ,
\quad \xi^{\nu}_{\mu}=\delta^{\nu}_{\mu}-n_{\mu}n^{\nu}\; ,
\eea 
where Greek indices are five dimensional and 
$n_{\mu}$ is the unit vector normal to the boundary: 
$n_{\mu}=(l,0,0,0,0)$.   
The coefficient of $S_1$ can be determined from the condition of the
finiteness of the action (but in a less strict way than in AdS/CFT). 
Moreover, using the holographic renormailzation group method  
\cite{BVV}  
this coefficient can be determined uniquely 
 \cite{dscftNO}. 

We consider 5-dimensional Euclidean de Sitter space which
is identical with 5-dimensional sphere $S^{5}$:  
\bea
\label{metricS5}
ds^{2}_{{\rm S}_5}&=&l^2 \left( dy^{2} 
+ \gamma_{ij} dx^{i}dx^{j} \right) \; \nn
\gamma _{ij} &=& \sin^{2}y\; \hat{\gamma}_{ij} \; ,
\eea
where $\hat{\gamma_{ij}}dx^{i}dx^{j}$ 
describes the metric of S$_{4}$ with unit radius. 

The metric of ${\rm S}_4$ with the unit radius is given by
\be
\label{S4metric1}
d\Omega^2_4= d \chi^2 + \sin^2 \chi d\Omega^2_3\ .
\ee
Here $d\Omega^2_3$ is described by the metric of 3 dimensional 
unit sphere. If one changes the coordinate $\chi$ to 
$\sigma$ by $\sin\chi = \pm {1 \over \cosh \sigma}$, 
one obtains\footnote{
If we Wick-rotate the metric by $\sigma\rightarrow it$, we 
obtain the metric of de Sitter space:
\[
d\Omega^2_4\rightarrow ds_{\rm dS}^2
= {1 \over \cos^2 t}\left(-dt^2 + d\Omega^2_3\right)\ .
\]
}
\be
\label{S4metric2}
d\Omega^2_4= {1 \over \cosh^2 \sigma}\left(d \sigma^2 
+ d\Omega^2_3\right)\ .
\ee
Now one assumes 
the metric of 5 dimensional spacetime as follows:
\be
\label{metric1}
ds^2=l^2 \left(dy^2 + \e^{2A(y,\sigma)}
\tilde \gamma_{ij}dx^i dx^j\right)\ ,
\quad \tilde \gamma_{ij}dx^i dx^j \equiv d \sigma^2 
+ d\Omega^2_3\ .
\ee
Due to the assumption (\ref{metric1}), the actions  
(\ref{action}) have the following forms:
\bea
\label{actions2}
&& \SEH= -{l^3 V_3 \over 16\pi G}\int dy d\sigma \left\{\left( -8 
\partial_y^2 A - 20 (\partial_y A)^2\right){\e^{4A} 
\over l^2}\right. \nn
&& \qquad \left. +{1 \over l^2}\left(-6\partial_\sigma^2 A 
 - 6 (\partial_\sigma A)^2 
+ 6 \right)\e^{2A} - {12 \over l^2} \e^{4A}\right\} \nn
&& \SGH= -{l^3 V_3 \over 2\pi G}\int d\sigma \e^{4A} 
\partial_y A \ ,\quad 
S_1= {3l^3 V_3 \over 8\pi G}\int d\sigma \e^{4A} \ .
\eea
Here $V_3$ is the volume or area of the unit 3 sphere
and it is assumed that there is a brane at $y=$constant.

In the bulk, one obtains the following equation of motion 
from $\SEH$ by the variation over $A$:
\be
\label{eq1a}
0= \left(-24 \partial_y^2 A - 48 (\partial_y A)^2 - 48 
\right)\e^{4A} + \left(-12 \partial_\sigma^2 A 
- 12 (\partial_\sigma A)^2 + 12\right)\e^{2A}\ ,
\ee
which corresponds to one of the Einstein equations. 
Then one finds solutions, $A_S$, which correspond to 
the metric  dS$_5$: 
\be
\label{blksl}
A=A_S=\ln\sin y  - \ln \cosh\sigma\ .
\ee

On the brane at the boundary, 
one gets the following equation:
\be
\label{eq2a}
0={48 l^3 \over 16\pi G}\left(\partial_y A - 1 \right)\e^{4A} \ .
\ee
Substituting the bulk solution $A=A_S$ in (\ref{blksl}) into 
(\ref{eq2a}) and defining the radius $R_b$ of the brane by
$R_b\equiv l\sin y_0 $, one obtains
\be
\label{slbr2}
0={1 \over \pi G}\left({1 \over R_b}\sqrt{1 - {R_b^2 \over l^2}}
 - {1 \over l}\right)R_b^4 \ .
\ee
First we should note $0\leq R_b\leq l$ by  definition. 
Eq.(\ref{slbr2}) has a solution\cite{SN}:
\be
\label{Csol}
R_b^2=R_{b0}^2\equiv 
{l^2 \over 2}\ \mbox{or}\ y_0 ={\pi \over 4},\ {3\pi \over 4}\ .
\ee
In Eq.(\ref{slbr2}), the first term 
${R_b^3 \over \pi G}\sqrt{1 - {R_b^2 \over l^2}}$ 
corresponds to the gravity, which makes the radius $R_b$ larger. 
On the other hand, the second term 
$-{R_b^4 \over \pi Gl}$ corresponds to the tension, which makes 
$R_b$ smaller. When $R_b<R_{b0}$,  gravity becomes larger than the 
tension and vice versa when $R_b>R_{b0}$. 
Then both of the solutions  
(\ref{Csol}) are stable. Although it is not clear from 
(\ref{slbr2}), $R_b=l$ ($y ={\pi \over 2}$) corresponds 
to the local maximum. 
This is because the point $y ={\pi \over 2}$ is a symmetric 
point. At the point, since the strength of  gravitational forces 
from two bulk spaces is equal and their signs are different 
with each other, the gravitaional forces to the brane 
are balanced. Furthermore, since $R_b$ becomes maximum there, 
the energy coming from the tension of the brane is maximum. 
Then the point $y ={\pi \over 2}$ becomes a local maximum for both of 
the gravitational force and tension of the brane. 
To see the above situation more, clearly we consider an effective 
potential $V(y)$. Since Eq.(\ref{eq2a}) are obtained from 
the variation of $A$, maybe we can define $V(y)$ as follows,
\bea
\label{pot1}
V(y)&={1 \over \pi G}&\int \left({1 \over R_b}
\sqrt{1 - {R_b^2 \over l^2}} - {1 \over l}\right)R_b^4 
\left.dA\right|_{\sigma:{\rm fixed}} \nn
&=&{l^3 \over \pi G} \int \left({\left|\cos y \right| \over 
\sin y} - 1\right)\sin^4y\cdot {\cos y \over \sin y}dy \nn
&=& V_0 + {l^3 \over \pi G}\left\{ {1 \over 8}\left(\sin y 
 - 2\sin^3 y\right)\left| \cos y \right| + {1 \over 8}
 \left| y - {\pi \over 2}\right| + {1 \over 4}\sin^4 y\right\} \ .
\eea
Here we used Eq.(\ref{blksl}) and $V_0$ is a constant of 
the integration. Then we find
\bea
\label{pot2}
&& V(0) - V_0 = V(\pi) - V_0 
= {l^3 \over \pi G}{\pi \over 16}\ ,\quad
V\left({\pi \over 4}\right) - V_0 
= V\left({3\pi \over 4}\right) - V_0 =
 {l^3 \over \pi G}\left({1 \over 16} 
 - {\pi \over 32}\right)\ ,\nn 
&& V\left({\pi \over 2}\right) - V_0 
= {l^3 \over \pi G}{1 \over 4}\ ,
\quad V'(0)=V'\left({\pi \over 4}\right)
=V'\left({\pi \over 2}\right)
=V'\left({3\pi \over 4}\right)=V'(\pi)=0 \ .
\eea
The conceptual (not exact) shape of the potential is 
given in Fig.\ref{Fig1}. From the figure, we can find that there are 
local maxima at $y=0$, ${\pi \over 2}$ and $\pi$ and minima at 
$y={\pi \over 4}$ and ${3\pi \over 4}$.

\unitlength 0.9mm
\begin{figure}
\begin{picture}(150,120)
\thicklines
\put(30,10){\vector(1,0){110}}
\put(30,10){\vector(0,1){110}}
\put(22,122){$V(y)- V_0$}
\put(142,8){$y$}
\qbezier(30,90)(34,90)(42,82)
\qbezier(42,82)(50,74)(55,74)
\qbezier(55,74)(59,74)(67,92)
\qbezier(67,92)(75,110)(80,110)
\qbezier(80,110)(85,110)(93,92)
\qbezier(93,92)(101,74)(105,74)
\qbezier(105,74)(110,74)(118,82)
\qbezier(118,82)(126,90)(130,90)
\put(5,73){${l^3 \over \pi G}\left({1 \over 16} 
 - {\pi \over 32}\right)$}
\put(18,89){${l^3 \over \pi G}{\pi \over 16}$}
\put(20,109){${l^3 \over \pi G}{1 \over 4}$}
\put(27,7){$0$}
\put(53,5){${\pi \over 4}$}
\put(78,5){${\pi \over 2}$}
\put(103,5){${3\pi \over 4}$}
\put(128,5){$\pi$}
\thinlines
\put(30,90){\line(1,0){100}}
\put(130,10){\line(0,1){80}}
\put(30,74){\line(1,0){75}}
\put(55,10){\line(0,1){64}}
\put(105,10){\line(0,1){64}}
\put(30,110){\line(1,0){50}}
\put(80,10){\line(0,1){100}}
\end{picture}
\caption{Conceptual shape of the effective potential $V(y)$ 
with respect to the position of the brane $y$.}
\label{Fig1}
\end{figure}
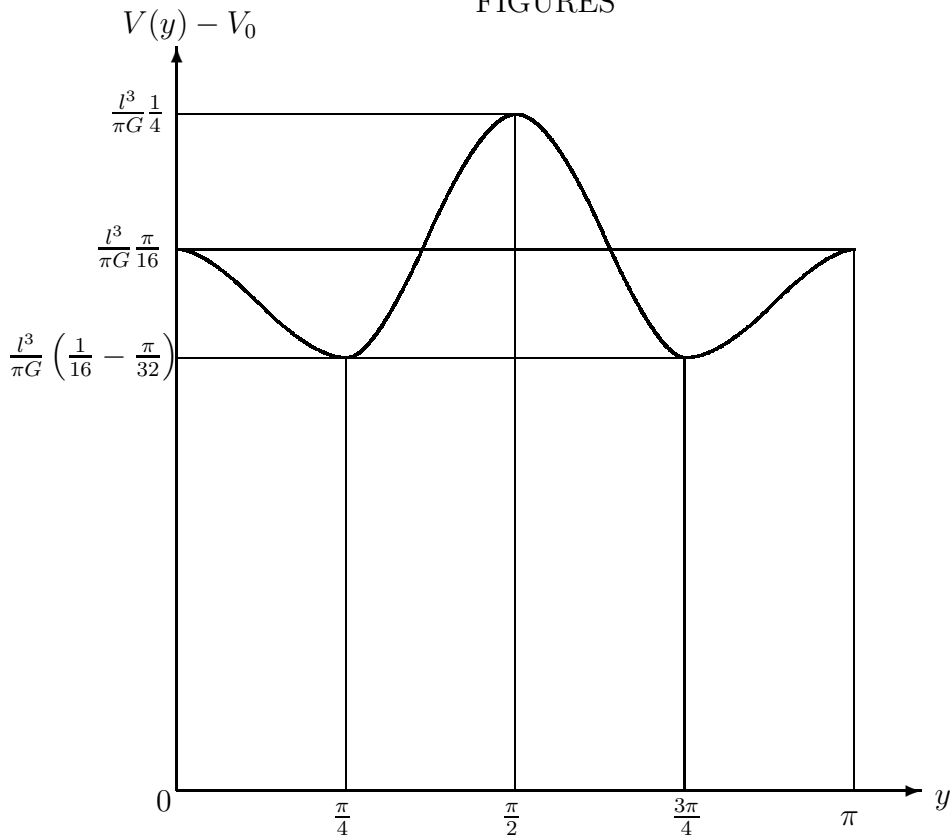

Hence, the possibility of creation of inflationary brane in 
de Sitter bulk is possible already on classical level, even 
in situation when brane tension is fixed by holographic RG. 
That is qualitatively different from the case of AdS bulk 
where only quantum effects led to creation of inflationary 
brane \cite{NOZ,HHR} (when brane tension was not free 
parameter).

It is interesting to study the effect of the perturbations 
of the gravity on the sphere. In case when bulk is AdS space it has been 
examined in ref.\cite{HHR} with the help of the AdS/CFT correspondence. 
We also study it applying the considerations from the proposed 
dS/CFT correspondence (for the introduction, see \cite{strominger}).
 Then we rewrite bulk metric which includes the
effect of the perturbation $h_{ij}(y,x)$ as follows:
\bea
\label{met}
ds^{2}_{{\rm S}_5}=l^2 dy^{2} + \left( l^2 \sin^{2}y\; \hat{\gamma_{ij}}+ 
h_{ij}(y,x) \right) dx^{i}dx^{j}  \; .
\eea  
$h_{ij}$ is transverse and traceless with respect to the metric
on the spherical spatial sections:
\bea
\label{eq1}
\hat{\gamma}^{ij}h_{ij}(y,x)=\hat{\nabla}^{i}h_{ij}(y,x)=0,
\eea
with $\hat{\nabla}$ being the covariant derivative
defined by the metric $\hat{\gamma}_{ij}$.  
We use the Greek letters $\mu$, $\nu$, $\cdots$ for 5 dimensional 
vector indeces and the Roman letters $i$, $j$, $\cdots$ for 
4 dimensional indeces. 
The Einstein equation in the bulk leads to the
classical solutions of scalar curvature $R$, 
Ricci tensor $R_{\mu\nu}$ and Riemann tensor 
$R_{\mu\rho\nu\sigma}$ as follows:
\bea
\label{eq2}
R&=&{D \over D-2}\Lambda \; ,\quad 
R_{\mu\nu}={\Lambda \over D-2}g_{\mu\nu}\; ,\nn
R_{\mu\rho\nu\sigma} &=& {\Lambda \over (D-1)(D-2)}
\left( g_{\mu\nu} g_{\rho\sigma} - g_{\mu\sigma}g_{\rho\nu} \right)\; .
\eea
Here $D$ and $\Lambda$ is the bulk dimension and cosmological 
constant respectively. We take $D=5$ and $\Lambda$ is related 
with the radius of S$_5$ by $\Lambda = {12 \over l^2}$.  
In the background of S$_4$, the linearized Einstein equation has 
the following form:
\bea
\label{eq3}
\nabla^{2}h_{\mu\nu}={2\over l^2}h_{\mu\nu}\; .
\eea
We now use the standard tensor spherical harmonics $H_{ij}^{(p)}(x)$ to
expand the metric perturbation
\bea
\hat{\gamma}^{ij}H^{(p)}_{ij}(x)=\hat{\nabla}^{i}H_{ij}^{(p)}(x)=0,
\eea  
 They are tensor eigenfunctions of the Laplacian:
\bea
\label{lapl}
\hat{\nabla}^2 H_{ij}^{(p)} =(2-p(p+d-1)) H_{ij}^{(p)} \; ,
\eea
where $p=2,3,\cdots$, and $d$ is the dimension of the boundary, 
$D=d+1$. As usual in the quantum mechanics, the eigenvalue of 
$-\nabla^2$ can be regarded as the square of the momentum. 
Especially when $p$ is large, we can regard $p$ as the 
absolute value of the momentum. 
The metric perturbation can be written as a sum of 
separable perturbations of the form
\bea
h_{ij}(y,x)=f_{p}(y)H_{ij}^{(p)}(x) \; .
\eea
Substituting this into (\ref{eq3}), we obtain the 
following differential equation:
\bea
&&f''_{p}(y) + (d-4)\cot y \; f'_{p}(y) \nn
&& -\left\{ -2(d-2)+ \left[ p(p+d-1)+2(d-3) 
\right] \sin ^{-2} y \right \} f_{p}(y) = 0 \;. 
\eea
Here $'$ denotes the derivative with respect to $y$. 
The solution can be written by associated Legendre functions:
\bea
\label{sol}
f_{p}(y) \propto (\sin y)^{(5-d)/2}
P^{-(p+(d-1)/2)}_{-(d+1)/2}( \cos y) \; .
\eea
The full solution for the metric perturbation is
\bea
\label{hij}
h_{ij}(y,x) = \sum_{p}{f_{p}(y) \over f_{p}(y_{0})}H_{ij}^{p}(x)
\int d^{4} x' \sqrt{\hat{\gamma} } h^{kl} (x') H_{kl}^{(p)}(x')
\; .
\eea
Here $h^{kl} (x)$ is the boundary value of the metric 
perturbation. This discussion generalizes the study of ref.\cite{HHR} 
done in AdS bulk to dS bulk.

By using equations (\ref{eq1}),(\ref{eq2}),(\ref{eq3}),  
one can obtain the explicit form of the action (\ref{action}) 
from the perturbated metric (\ref{met}).
Then the Einstein-Hilbert action with the perturbations is 
given by 
\bea
S_{\rm EH} &=& -{1\over 16\pi G} \int d^{5}x \sqrt{ g }  
\left( R -{12\over l^2}\right)\; , \nn
&=& -{1\over 16\pi G} \int d^{5}x \sqrt{ g }  
\left( {8\over l^2}+{1\over 4}h^{\mu\nu} \nabla ^2 h_{\mu\nu}
-{1\over 2 l^2 }h^{\mu\nu}h_{\mu\nu} \right)\; , \nn
&& -{1\over 16\pi G} \int d^{4}x \sqrt{\gamma }  
\left( -{1\over 2}n^{\mu}h^{\nu\rho}\nabla_{\nu}h_{\mu\rho}
+{3\over 4}h_{\nu\rho} n^{\mu}\nabla_{\mu} h^{\nu\rho} \right)\; . 
\eea 
Since the bulk part of the action always vanishes, the action 
is to be expressed in terms of the boundary values of the 
perturbation $h_{ij}$:
\bea
S_{\rm EH} &=&  -{l^3 \over 2\pi G} \int d^{4}x \sqrt{ \hat{\gamma} }  
\int^{y_{0}}_{0} dy \sin ^{4} y \nn
&& -{l^3 \over 16\pi G} \int d^{4}x \sqrt{ \hat{\gamma} }  
\left( {3\over 4 l^4} h^{ij}\partial_{y} h_{ij}
-{\cot y_{0} \over l^4}h^{ij}h_{ij} \right),
\eea
where we are now raising and lowering the indeces $i$, $j$, 
$\cdots$ by using $\hat{\gamma}_{ij}$ as in ref.\cite{HHR}. 
The Gibbons-Hawking term has the following form:
\bea
S_{\rm GH} &=& -{1\over 8\pi G} \int d^{4}x \sqrt{\gamma} 
\; K\; , \nn
&=& -{l^3 \over 2\pi G} \int d^{4}x \sqrt{\hat{\gamma}}
\left( \sin^{3}y_{0} \cos y_{0}-{1 \over 8 l^4}h^{ij}\partial_{y} 
h_{ij} \right)\; .  
\eea
The counter term $S_1$ is
\bea
S_{1} &=& {3\over 8\pi G l}\int d^{4}x\sqrt{\gamma} \; , \nn
&=& {3 l^{3} \over 8\pi G}\int d^{4}x\sqrt{\hat{\gamma}}
\left( \sin^4 y_{0} - {1\over 4 l^4 } h^{ij}h_{ij} \right) \; .
\eea
Then the total action becomes
\bea
\label{total}
S &=& S_{\rm EH}+S_{\rm GH}+S_{1} \; ,\nn
&=& {l^{3} \over 2\pi G} \int d^{4}x \sqrt{\hat{\gamma}}
\left\{ -{3\over 4}\sin^{3}y_{0}\cos y_{0}-{3\over 8} y_{0} 
+{3\over 8}\cos y_{0}\sin y_{0} \right. \nn
&& \left. +{1 \over 32 l^4} h^{ij}\partial_{y} h_{ij} + {1\over 16 l^{4}}
(-3+2\cot y_{0} ) h^{ij}h_{ij} +{3\over 4} \sin^4 y_{0} \right\} \; .
\eea
Eq.(\ref{total}) describes the fluctuation of S$_4$ brane in the 
bulk S$_5$, or after Wick-rotation, 3 dimensional dS brane in the 
bulk 5-dimensional dS space.

In case that the bulk space is 5 dimensional Euclidean 
anti-de Sitter space (hyperboloid), if we choose the metric as 
\bea
\label{metricH5}
ds^{2}_{{\rm S}_5}&=&l^2 \left( dy^{2} 
+ \gamma_{ij} dx^{i}dx^{j} \right) \; \nn
\gamma _{ij} &=& \sinh^{2}y\; \hat{\gamma}_{ij} \; ,
\eea
instead of (\ref{metricS5}) and assume that the brane lies at 
$y=y_0$, the equation corresponding to (\ref{total} has the 
following form
\bea
\label{totalH}
S^{\rm AdS} &=& {l^{3} \over 2\pi G} \int d^{4}x \sqrt{\hat{\gamma}}
\left\{ -{3\over 4}\sinh^{3}y_{0}\cosh y_{0}-{3\over 8} y_{0} 
+{3\over 8}\cosh y_{0}\sinh y_{0} \right. \nn
&& \left. +{1 \over 32 l^4} h^{ij}\partial_{y} h_{ij} 
+ {1\over 16 l^{4}}
(-3+2\coth y_{0} ) h^{ij}h_{ij} +{3\over 4} 
\sinh^4 y_{0} \right\} \; .
\eea
The expression (\ref{totalH}) can be also obtained from 
(\ref{total}) by replacing $\sin$ and $\cos$ by $\sinh$ and 
$\cosh$, respectively. In fact, the function $f_p$ corresponding 
to (\ref{sol}) is also given by \cite{HHR},
\bea
\label{solH}
f^{\rm AdS}_{p}(y) \propto (\sinh y)^{(5-d)/2}
P^{-(p+(d-1)/2)}_{-(d+1)/2}( \cosh y) \; .
\eea

Now one can consider the Euclidean graviton correlator, 
$\left< h_{ij}(x)h_{i'j'}(x') \right>$ (for earlier study on 
four-dimensional graviton correlator in maximally symmetric spaces, see
\cite{graviton} and references therein).  To 
read off graviton correlator from eq.(\ref{total}), we focus on the
$h^{ij}\partial_{y} h_{ij}$ and $h^{ij}h_{ij}$ terms.  The 
expansion of $\partial_{y} h_{ij}$ at $y=y_0$ is obtained by
using (\ref{hij}) as
\bea
\partial_{y} h_{ij}= \sum_{p}{f'_{p}(y_0) \over f_{p}(y_{0})}H_{ij}^{p}(x)
\int d^{4} x' \sqrt{\hat{\gamma} } h^{kl} (x') H_{kl}^{(p)}(x')\; .
\eea
The terms related with the graviton correlator in the action (\ref{total})
can be rewritten as
\bea
\label{4dcol}
S_{\rm correlator} = \sum_{p} 
\left( \int d^{4} x' \sqrt{\hat{\gamma} } h^{kl} (x') H_{kl}^{(p)}(x')
\right)^2 \; T_{p}(y_0) \; , 
\eea
where
\bea
\label{tp5}
T_{p}(y_0) = {l^3 \over 2\pi G} {1\over 32 l^4}
\left( {f'_{p}(y_0) \over f_{p}(y_0) }+ 4 \cot y_0 - 6 \right) \; .
\eea
With above equations one arrives at the graviton correlator as
\bea
\label{bbcrltr}
\left< h_{ij}(x)h_{i'j'}(x') \right> &=& \sum_{p=2}^{\infty}
W_{iji'j'}^{(p)} (x,x') {1\over 2 T_{p}(y_0)}\; , \nn
W_{iji'j'}^{(p)} (x,x') &\equiv& H^{(p)}_{ij}(x)H^{(p)}_{i'j'}(x')\; .
\eea
Using (\ref{sol}), we can rewrite (\ref{tp5}) for $d=4$ as 
following form,
\bea
\label{ind0}
T_{p}(y_0)= {1 \over 64\pi Gl}\left\{
 - 6 -\left( p- 3 \right) \cot y_{0} 
 -{P^{-p-1/2}_{-5/2} (\cos y_{0}) \over P^{-p-3/2}_{-5/2} 
(\cos y_{0})}\right\}\; . 
\eea
For our solutions $y_{0}={\pi \over 4},{3 \pi \over 4}$  
(\ref{Csol}), $T_{p}(y_0)$ becomes
\bea
\label{ind1}
T_{p}\left( {\pi \over 4} \right) &=& 
{1 \over 64\pi Gl}\left\{ - p - 3  
 -{P^{-p-1/2}_{-5/2}({1\over \sqrt{2}}) \over P^{-p-3/2}_{-5/2} 
({1\over \sqrt{2}} ) }\right\}\; , \nn 
T_{p}\left({3 \pi \over 4} \right) &=& {1 \over 64\pi Gl}\left\{
 p - 9 -{P^{-p-1/2}_{-5/2}(-{1\over \sqrt{2}}) \over 
P^{-p-3/2}_{-5/2} (-{1\over \sqrt{2}} ) }\right\}\; . 
\eea

Now one can consider the perturbation based on 4-dimensional action, 
in order to compare it with that of the 4-dimensional gravity 
induced from the bulk 5-dimensional dS gravity. 
We now introduce 4-dimensional cosmological constant 
$\Lambda^{(4)}$ and the Newton constant $G^{(4)}$, which 
would be adjusted later to be 
consistent with the induced gravity.  The perturbed 
4-dimensional Einstein-Hilbert action is given as follows:
\bea
\label{actionEin}
S^{(4)}_{\rm EH} &=& -{1\over 16 \pi G^{(4)}} \int d^4 x 
\sqrt{g} \left(R+\Lambda ^{(4)} \right)\; , \\
&=& S_{0}^{(4)}  -{1\over 16 \pi G^{(4)}} \int d^4 x \sqrt{g}
\left\{ {1\over 4} h^{ij} \hat{\nabla}^2 h_{ij} 
+\left( -{1\over 4}\Lambda ^{(4)}-{2 \over l^2 \sin^{2} y_{0} } 
\right) h^{ij}h_{ij} \right\} \; \nonumber .  
\eea
Using eq.(\ref{lapl}), we can obtain the
function for calculating graviton correlator 
(instead of (\ref{tp5})) as 
\bea
\label{Ein1}
T^{(4)}_{p}(y_0) = -{1 \over 16\pi G^{(4)}}\left\{
{1\over 4l^2\sin^2 y_0}(2- p(p+3) ) 
-{1\over 4}\Lambda ^{(4)}-{2 \over l^2 \sin^{2} y_{0} }\right\} \; . 
\eea

Eq.(\ref{lapl}) tells that $p$ ($=2$, $3$, $\cdots$) is 
proportional to the magnitude of the momentum of the graviton. 
In the analogy with the Randall-Sundrum model \cite{RS}, 
we expect the present model of the 4 dimensional gravity induced 
from the 5 dimensional bulk gravity would reproduce the 4-dimensional 
Einstein gravity in the low-energy limit. Then we compare 
(\ref{ind1}) and (\ref{Ein1}) for $p=2$, $3$ cases and determine 
$G^{(4)}$ and $\Lambda^{(4)}$ so that the two expressions 
coincide with each other. 
First we consider $y_0={\pi \over 4}$ solution. Then one has
\be
\label{Ein2}
{1 \over 64\pi G l}\left( -5 - 
{P^{-{5 \over 2}}_{-{5 \over 2}}\left({1 \over \sqrt{2}}\right) \over 
P^{-{7 \over 2}}_{-{5 \over 2}}\left({1 \over \sqrt{2}}\right)}
\right)
= -{1 \over 16\pi G^{(4)}}\left(-{8 \over l^2} 
- {\Lambda^{(4)} \over 4}\right)
\ee
for $p=2$ and 
\be
\label{Ein3}
{1 \over 64\pi G l}\left( -6 - 
{P^{-{7 \over 2}}_{-{5 \over 2}}\left({1 \over \sqrt{2}}\right) \over 
P^{-{9 \over 2}}_{-{5 \over 2}}\left({1 \over \sqrt{2}}\right)}
\right)
= -{1 \over 16\pi G^{(4)}}\left(-{12 \over l^2} 
- {\Lambda^{(4)} \over 4}\right)
\ee
for $p=3$. As a result
\bea
\label{Ein4}
\Lambda^{(4)}&=&{16 \over l^2}
{3 + 3 {P^{-{5 \over 2}}_{-{5 \over 2}}\left({1 \over \sqrt{2}}\right) 
\over P^{-{7 \over 2}}_{-{5 \over 2}}\left({1 \over \sqrt{2}}\right)}
 -2 {P^{-{7 \over 2}}_{-{5 \over 2}}\left({1 \over \sqrt{2}}\right) \over 
P^{-{9 \over 2}}_{-{5 \over 2}}\left({1 \over \sqrt{2}}\right)} \over 
1 - {P^{-{5 \over 2}}_{-{5 \over 2}}\left({1 \over \sqrt{2}}\right) 
\over P^{-{7 \over 2}}_{-{5 \over 2}}\left({1 \over \sqrt{2}}\right)}
+ {P^{-{7 \over 2}}_{-{5 \over 2}}\left({1 \over \sqrt{2}}\right) \over 
P^{-{9 \over 2}}_{-{5 \over 2}}\left({1 \over \sqrt{2}}\right)} } \nn
{1 \over 16\pi G^{(4)}}&=&-{l \over 256\pi G}
\left(1 - {P^{-{5 \over 2}}_{-{5 \over 2}}\left({1 \over \sqrt{2}}\right) 
\over P^{-{7 \over 2}}_{-{5 \over 2}}\left({1 \over \sqrt{2}}\right)}
+ {P^{-{7 \over 2}}_{-{5 \over 2}}\left({1 \over \sqrt{2}}\right) \over 
P^{-{9 \over 2}}_{-{5 \over 2}}\left({1 \over \sqrt{2}}\right)}
\right)\ .
\eea
This gives the 4d parameters $G^{(4)}$ and $\Lambda^{(4)}$ 
in terms of the 5d parameters so that 
$T_{p}\left( {\pi \over 4} \right) $ coincides with $T^{(4)}_p$ 
for $p=2$ and $3$, which corresponds to the infrared behavior. 
Of course, there will appear some difference for higher $p=4$, 
$5$, $6$, $\cdots$. As we will see later, with account of 
the quantum effects from the matter fields on the brane, the 
difference between $T_{p}\left( {\pi \over 4} \right) $ and 
$T^{(4)}_p$ disappears in the ultraviolet region. Then  both 
of infrared and ultraviolet behaviors can be reproduced from the 
induced gravity.  

On the other hand, for the solution $y_0={3\pi \over 4}$, one gets
\be
\label{Ein5}
{1 \over 64\pi G l}\left( -7 - 
{P^{-{5 \over 2}}_{-{5 \over 2}}\left(-{1 \over \sqrt{2}}\right) \over 
P^{-{7 \over 2}}_{-{5 \over 2}}\left(-{1 \over \sqrt{2}}\right)}
\right)
= -{1 \over 16\pi G^{(4)}}\left(-{8 \over l^2} 
- {\Lambda^{(4)} \over 4}\right)
\ee
for $p=2$ and 
\be
\label{Ein6}
{1 \over 64\pi G l}\left( -6 - 
{P^{-{7 \over 2}}_{-{5 \over 2}}\left(-{1 \over \sqrt{2}}\right) \over 
P^{-{9 \over 2}}_{-{5 \over 2}}\left(-{1 \over \sqrt{2}}\right)}
\right)
= -{1 \over 16\pi G^{(4)}}\left(-{12 \over l^2} 
- {\Lambda^{(4)} \over 4}\right)
\ee
for $p=3$. Then 
\bea
\label{Ein7}
\Lambda^{(4)}&=&-{16 \over l^2}
{9 + 3 {P^{-{5 \over 2}}_{-{5 \over 2}}\left(-{1 \over \sqrt{2}}\right) 
\over P^{-{7 \over 2}}_{-{5 \over 2}}\left(-{1 \over \sqrt{2}}\right)}
 -2 {P^{-{7 \over 2}}_{-{5 \over 2}}\left(-{1 \over \sqrt{2}}\right) \over 
P^{-{9 \over 2}}_{-{5 \over 2}}\left(-{1 \over \sqrt{2}}\right)} \over 
1 + {P^{-{5 \over 2}}_{-{5 \over 2}}\left(-{1 \over \sqrt{2}}\right) 
\over P^{-{7 \over 2}}_{-{5 \over 2}}\left(-{1 \over \sqrt{2}}\right)}
- {P^{-{7 \over 2}}_{-{5 \over 2}}\left(-{1 \over \sqrt{2}}\right) \over 
P^{-{9 \over 2}}_{-{5 \over 2}}\left(-{1 \over \sqrt{2}}\right)} } \nn
{1 \over 16\pi G^{(4)}}&=&{l \over 256\pi G}
\left(1 + {P^{-{5 \over 2}}_{-{5 \over 2}}\left(-{1 \over \sqrt{2}}\right) 
\over P^{-{7 \over 2}}_{-{5 \over 2}}\left({-1 \over \sqrt{2}}\right)}
- {P^{-{7 \over 2}}_{-{5 \over 2}}\left(-{1 \over \sqrt{2}}\right) \over 
P^{-{9 \over 2}}_{-{5 \over 2}}\left(-{1 \over \sqrt{2}}\right)}
\right)\ .
\eea
Again, the effective Newton constant and the effective
cosmological constant of the induced 4-dimensional gravity are expressed 
in terms of the bulk 5-dimensional gravity parameters. 

We now investigate the short distance behaviour, which is determined 
from the large $p$ behavior. When $p$ is large
\be
\label{Ein8}
P^{-p-{1 \over 2}}_{-{5 \over 2}}\left(\cos y_0\right)
\rightarrow {\e^{\left(-p - {1 \over 2}\right)\pi i}
\over \Gamma\left({3 \over 2} + p\right)}
\left({1 + \cos y_0 \over 1 - \cos y_0}\right)^{-{p + {1 \over 2} 
\over 2}}\ .
\ee
Then one obtains
\be
\label{Ein9}
{P^{-p-{1 \over 2}}_{-{5 \over 2}}\left(\cos y_0\right)
\over P^{-p-{3 \over 2}}_{-{5 \over 2}}\left(\cos y_0\right)}
\rightarrow - \left({3 \over 2}+p\right)\sqrt{{1+\cos y_0 \over 
1 - \cos y_0}}\ .
\ee
Especially for the solution $y_0={\pi \over 4}$, we find
\be
\label{Ein10}
{P^{-p-{1 \over 2}}_{-{5 \over 2}}\left({1 \over \sqrt{2}}\right)
\over P^{-p-{3 \over 2}}_{-{5 \over 2}}\left({1 \over \sqrt{2}}
\right)}\rightarrow - \left({3 \over 2} + p\right)
\left(\sqrt{2} + 1\right)
\ee
and for the solution $y_0={3\pi \over 4}$, one gets
\be
\label{Ein11}
{P^{-p-{1 \over 2}}_{-{5 \over 2}}\left(-{1 \over \sqrt{2}}\right)
\over P^{-p-{3 \over 2}}_{-{5 \over 2}}\left(-{1 \over \sqrt{2}}
\right)}\rightarrow - \left({3 \over 2} + p\right)
\left(\sqrt{2} - 1\right)\ .
\ee
Then $T_p$  (\ref{ind1}) is linear for large $p$:
\be
\label{Ein12}
T_{p}\left( {\pi \over 4} \right) \rightarrow 
-{\left(\sqrt{2} + 2\right)p \over 64\pi Gl}\; ,\quad 
T_{p}\left({3 \pi \over 4} \right) \rightarrow 
-{\left(\sqrt{2} -2\right)p \over 64\pi Gl}\; .
\ee
On the other hand, $T^{(4)}_p$ (\ref{Ein1}) is proportional to 
the square of $p$:
\be
\label{Ein13}
T^{(4)}_{p}(y_0) \rightarrow {p^2 \over 64\pi G^{(4)}l^2\sin^2y_0} \; . 
\ee
Then the short distance behavior seems to be changed drastically. 
Eq.(\ref{Ein13}) from the pure 4d theory tells that the 
graviton correlator behaves as 
\be
\label{Cr1}
\left< h_{ij}(x)h_{i'j'}(x') \right> 
\sim \int {d^4p \over p^2}\e^{ip\cdot\left(x-x'\right)}
\sim {1 \over \left|x-x'\right|^2}\ ,
\ee
which is consistent with the Coulomb-like behavior in 4 
dimensions. On the other hand, Eq.(\ref{Ein12}) gives
\be
\label{Cr2}
\left< h_{ij}(x)h_{i'j'}(x') \right> 
\sim \int {d^4p \over p}\e^{ip\cdot\left(x-x'\right)}
\sim {1 \over \left|x-x'\right|^3}\ .
\ee
The behavior of $\left|x-x'\right|^{-3}$ is nothing but the 
5-dimensional one. Therefore the potential of the gravity 
behaves as ${1 \over r^2}$ rather than ${1 \over r}$. Then the gravity 
force in the induced gravity becomes stronger at short 
distances than that in the pure 4d one. 
This indicates that one can observe the bulk 5 
dimensional space effects from the short distance behavior of 
the correlator (\ref{bbcrltr}). 

As a variation, we consider the case that the bulk space is 
anti-de Sitter space by replacing $l^2$ in $S_{\rm EH}$  
(\ref{action}). If we choose $S_1$ as in (\ref{action}), however, 
there is no solution describing the S$_4$ brane without adding 
the conformal anomaly induced action. Here as a toy model, we 
choose the tension of the brane to be arbitrary and replace $l$ by 
another parameter $l'$ in $S_1$. Then one starts from the 
following action:
\bea
\label{actionAdS}
S &=& S_{\rm EH}+S_{\rm GH}+S_{1} ,\quad 
S_{\rm EH} = -{1\over 16\pi G} \int d^{5}x \sqrt{ g }  
\left( R + {12\over l^2}\right)\; , \nn
S_{\rm GH} &=& -{1\over 8\pi G} \int d^{4}x \sqrt{\gamma} 
\; K \; , \quad S_{1}=
{3\over 8\pi G l'}\int d^{4}x\sqrt{\gamma} \; .
\eea
We now choose the metric of the bulk AdS (hyperboloid) as 
in (\ref{metricH5}). 
If there is a brane at $y=y_0$, by 
defining the radius of S$_4$ by $R_b=l\sinh y_0$, we obtain 
the equation corresponding to (\ref{slbr2}) in the following 
form:
\be
\label{slbr2H}
{1 \over R_b}\sqrt{1 + {R_b^2 \over l^2}}= {1 \over l'} \ .
\ee
When $R_b\to +\infty$, the r.h.s. of (\ref{slbr2H}) 
goes to ${1 \over l}$ and when $R_b\to 0$, the r.h.s. goes to the 
positive infinity. Then if 
\be
\label{HH1}
l'<l\ ,
\ee
there is a unique non-trivial solution. 
The equation corresponding to (\ref{tp5}) has the folowing form:
\bea
\label{tp5H}
T_{p}(y_0) = {l^3 \over 2\pi G} {1\over 32 l^4}
\left( {{f^{\rm AdS}}'_{p}(y_0) \over f^{\rm AdS}_{p}(y_0) }
+ 4 \coth y_0 - {6l \over l'} \right) \; .
\eea
Here $f^{\rm AdS}_{p}(y_0)$ is defined by (\ref{solH}). Then the 
large $p$-behavior is similar to (\ref{Ein12}):
\be
\label{tp6}
T_{p}(y_0) \to {p \over 64\pi Gl} {\cosh y_0 + 1 \over 
\sinh y_0}\ .
\ee
The potential of the gravity behaves as ${1 \over r^2}$ 
rather than ${1 \over r}$ at the short distance, again. Then the gravity 
force in the induced gravity becomes stronger at short 
distances than that in the pure 4d one. 
Thus, we demonstrated that qualitative behaviour of 4d induced gravity
obtained from dS or AdS bulk is similar, but it differs from
the standard 4d gravity behaviour.

Note that in order to obtain a behavior of the graviton correlator 
similar to 
\cite{HHR}, the radius of the brane should be much larger than 
the length parameter of the bulk de Sitter space. If we
consider the Euclidean bulk, the radius is always smaller than the
length parameter. Then one needs to  continue analytically the bulk
metric to Lorentzian.
The related interesting question is about the dependence of 
our results from the 
brane tension. Imagine that we do not fix the boundary counterterm and  
 start with the following action, instead of (\ref{action}), 
\bea
\label{actionL}
S &=& S_{\rm EH}+S_{\rm GH}+S_{1} ,\quad 
S_{\rm EH} = -{1\over 16\pi G} \int d^{5}x \sqrt{ -g }  
\left( R -{12\over l^2}\right)\; , \nn
S_{\rm GH} &=& -{1\over 8\pi G} \int d^{4}x \sqrt{\gamma} 
\;\; , \quad S_{1}= -
{3\over 8\pi G l'}\int d^{4}x\sqrt{\gamma} \ .
\eea
Here we choose the coefficient of $S_1$ to be arbitrary, by 
replacing $l$ in $S_1$  (\ref{action}) by $l'$.
Then we obtain the 5-dimensional Lorentzian de Sitter space:  
\bea
\label{metricS5L}
ds^{2}_{{\rm dS}_5}&=&l^2 \left( - dy^{2} 
+ \gamma_{ij} dx^{i}dx^{j} \right) \; \nn
\gamma _{ij} &=& \cosh^{2}y\; \hat{\gamma}_{ij} \; ,
\eea
which can be obtained from (\ref{metricS5}) by analytically 
continuing $y$ as
\be
\label{an}
y\to {\pi \over 2} + iy\ .
\ee
Then the equation 
corresponding to (\ref{eq2a}) has the following form:
\be
\label{L1}
\left.\partial_y A\right|_{\rm on the brane}= {l \over l'}\ .
\ee 
Here $A$ is given by
\be
\label{L2}
A=\ln \cosh y - \ln \cosh \sigma\ .
\ee
Defining the radius $R_b$ of the brane by $R_b=l\cosh y_0$, 
and assuming that there is a brane at $y=y_0$, Eq.(\ref{L1}) 
can be rewritten by
\be
\label{L3}
{1 \over R_b}\sqrt{{R_b^2 \over l^2}-1}={l \over l'}\ .
\ee
Eq.(\ref{L3}) has a solution labeled by $l'$ if ${l \over l'}<1$ and 
when ${l \over l'}\to 1 - 0$, $R_b$ goes to infinity. 

Then the graviton correlator may be calculated as following
\bea
\label{L4}
\left< h_{ij}(x)h_{i'j'}(x') \right> &=& \sum_{p=2}^{\infty}
W_{iji'j'}^{(p)} (x,x') {1\over 2 T_{p}(y_0)}\; , \nn
T_{p}(y_0) &=& -{l^3 \over 2\pi G} {1\over 32 l^4}
\left( {f'_{p}(y_0) \over f_{p}(y_0) }+ 4 \tanh y_0
 - {6l \over l'} \right) \; ,\nn
f_{p}(y) &\propto& (\cosh y)^{1/2}\left(
P^{-(p+3/2)}_{-5/2}( i\sinh y)
+ P^{-(p+3/2)}_{-5/2}( -i\sinh y)\right)\ .
\eea
We now consider the case that the radius of the brane $R_b$ is
large, then $y_0$ is also large and ${l \over l'}\to 1 - 0$.
Since when $R_b$ is large
\be
\label{L5}
{f'_{p}(y_0) \over f_{p}(y_0) }\to 2 - {l^2 \over 2R_b^2}
(p+1)(p+2) + {\cal O}\left({l^4 \over R_b^4}\right)\ ,
\ee
one gets
\be
\label{L6}
T_{p}(y_0) \to {l \over 128\pi G R_b^2}(p+1)(p+2)\ .
\ee
The corresponding equation in the pure 4d gravity is given in
(\ref{Ein1}) replacing $l\sin y_0$ by $R_b$:
\bea
\label{L7}
T^{(4)}_p(y_0) = -{1 \over 128\pi G^{(4)} R_b^2}\left\{
(2- p(p+3) ) - R_b^2 \Lambda^{(4)} - 8\right\} \; .
\eea
Then if we choose 
\be
\label{L8}
G^{(4)}={G \over l}\ ,\quad \Lambda^{(4)}=-{9 \over R_b^2}\ ,
\ee
we find $T_p(y_0)\to T^{(4)}_p(y_0)$ when the brane is large 
($R_b\to \infty$ or ${l \over l'}\to 1-0$). Thus, in this limit 
the induced gravity coincides with the pure 4d gravity where 
the Newton constant $G^{(4)}$ and the cosmological constant 
$\Lambda^{(4)}$ is given by (\ref{L8}). We should note again 
that the brane can exist only when $l'<l$. 

Note that in the correlation function 
$\left< h_{ij}(x)h_{i'j'}(x') \right>$ it is assumed 
$l,l'\ll \left|x-x'\right| \ll R_b$. That is, the distance 
$\left|x-x'\right|$ is much shorter than the radius of the 
brane $R_b$ but much larger than $l$ and $l'$. 
At very  
short distance $\left|x-x'\right|\ll l,l'$, the correlation 
function behaves as 5-dimensional one as in \ref{Cr2}. 

\section{Graviton correlator and  action with account of conformal 
quantum matter}

\newcommand\wlBox{\mbox{
\raisebox{0.1cm}{$\widetilde{\mbox{\raisebox{-0.1cm}\fbox{\ }}}$}}}
\newcommand\BBox{\mbox{\raisebox{0.1cm}\fbox{\ }}}
\newcommand\htBox{\mbox{
\raisebox{0.1cm}{$\hat{\mbox{\raisebox{-0.1cm}{$\Box$}}}$}}}

Let us try to analyze the effect of quantum CFT to graviton correlator.
As shown in (\ref{Ein12}) and (\ref{tp6}), $T_p(y_0)$ in the 
induced gravity behaves  linearly in $p$ for large $p$ and 
the graviton correlator shows its 5-dimensional structure at 
short distances. As pointed out in 
\cite{HHR,HHR2}, however, the short distance behavior is changed by 
including the quantum matter fields on the brane (by 
adding the effective action ${\cal W}$, which is induced from 
the conformal 
anomaly of the matter fields, to the action (\ref{action})): 
\bea
\label{Cr3}
&& {\cal W}= -b \int d^4x \sqrt{\widetilde \gamma}\widetilde F A 
 - b' \int d^4x \sqrt{\widetilde \gamma}
\left\{A \left[2 {\wlBox}^2 
+\widetilde R_{\mu\nu}\widetilde\nabla_\mu\widetilde\nabla_\nu 
 - {4 \over 3}\widetilde R \wlBox^2 \right.\right. \nn
&& \left.\left. \qquad 
+ {2 \over 3}(\widetilde\nabla^\mu \widetilde R)\widetilde\nabla_\mu
\right]A 
+ \left(\widetilde G - {2 \over 3}\wlBox \widetilde R
\right)A \right\} \nn
&& \qquad +{1 \over 12}\left\{b''+ {2 \over 3}(b + b')\right\}
\int d^4x \sqrt{\widetilde \gamma} \left[ \widetilde R - 6\wlBox A 
 - 6 (\widetilde\nabla_\mu A)(\widetilde \nabla^\mu A)
\right]^2 \ .
\eea 
Here one chooses the 4-dimensional boundary metric as 
\be
\label{tildeg}
\gamma_{ij}=\e^{2A}\tilde \gamma_{ij}
\ee 
and we specify the quantities with $\tilde \gamma_{ij}$ by 
using $\tilde{\ }$. 
$G$ ($\tilde G$) and $F$ ($\tilde F$) are the Gauss-Bonnet
invariant and the square of the Weyl tensor, which are given as
\be
\label{GF}
G=R^2 -4 R_{ij}R^{ij}
+ R_{ijkl}R^{ijkl} \ ,\quad 
F={1 \over 3}R^2 -2 R_{ij}R^{ij}
+ R_{ijkl}R^{ijkl} \ ,
\ee
In the effective action (\ref{Cr3}) induced by brane quantum 
matter, in general, with $N$ real scalar, $N_{1/2}$ 
Dirac spinor, $N_1$ vector 
fields, $N_2$  ($=0$ or $1$) gravitons and $N_{\rm HD}$ higher 
derivative conformal scalars, $b$, $b'$ and $b''$ are
\bea
\label{bs}
&& b={N +6N_{1/2}+12N_1 + 611 N_2 - 8N_{\rm HD} 
\over 120(4\pi)^2}\ ,\nn 
&& b'=-{N+11N_{1/2}+62N_1 + 1411 N_2 -28 N_{\rm HD} 
\over 360(4\pi)^2}\ ,\quad b''=0\ .
\eea
Usually, $b''$ may be changed by the finite renormalization 
of local counterterm in the gravitational effective action 
but this parameter plays an important role in tensor perturbations,
which leads to decay of de Sitter space (end of inflation) 
\cite{HHR2}. 

On the brane, we can choose $h^{i}_{i}=0$ as one of the gauge 
conditions. Then since the area of the brane is invariant under 
scale transformation, one can regard that the scale factor $A(x)$ 
is constant, $A(x)=A_0$ on the brane.
Under this condition, the eq.(\ref{Cr3}) becomes as
\bea
\label{wano}
&& {\cal W}= -b \int d^4x \sqrt{\widetilde \gamma}
\left( -{2\over 3}\widetilde R^2 + 2 \widetilde R_{ij} 
\widetilde R^{ij} \right) A_0 \nn
&& \qquad +{1 \over 12}\left\{b''+ {2 \over 3}(b + b')\right\}
\int d^4x \sqrt{\widetilde \gamma} \widetilde R^2 \; ,\nn
&& \qquad = -b \int d^4x \sqrt{\gamma}
\left( -{2\over 3}R^2 + 2 R_{ij} R^{ij}  \right) A_0 \nn
&& \qquad +{1 \over 12}\left\{b''+ {2 \over 3}(b + b')\right\}
\int d^4x \sqrt{\gamma} R^2 \; .
\eea
Here we use the relation $\sqrt{\widetilde \gamma}= e^{-4A_0}
\sqrt{\gamma}$ and $\widetilde R = e^{2A_0}R$, 
$\widetilde R_{ij}= R_{ij}$.
\bea
R &=& R_{(0)}+{1\over 2}\delta^2 R + {\cal O}\left(h^3\right) 
\; ,\nn
\delta^2 R &\equiv& {2 \over \tilde{l}^2 }h_{ij}h^{ij}
+{1\over 2}h^{ij}\nabla^2 h_{ij}+\nabla^2 (h_{ij}h^{ij})\; ,\nn
\sqrt{\gamma} &=& \sqrt{\gamma_{(0)}}\left( 1-{1\over 4}h^{ij}h_{ij} 
+ {\cal O}\left(h^3\right)\right) \; .
\eea
Here $\tilde l = R_b = l \sinh y_0$ ($\sin y_0$) when the bulk 5 
dimensional space is anti-deSitter (de Sitter). 
Then  neglecting $ {\cal O}\left(h^3\right)$ terms, one finds 
\bea
\label{rr2}
\int d^4x \sqrt{\gamma} R^2
&=& \int d^4x \sqrt{\gamma}
\left\{ R_{(0)}^2 -{1\over 4}h^{ij}h_{ij}R_{(0)}^2 \right. \nn
&& \quad \left. + \left( {2 \over \tilde{l}^2 }h_{ij}h^{ij}
+{1\over 2}h^{ij}\nabla^2 h_{ij}+\nabla^2 (h_{ij}h^{ij}) \right) 
R_{(0)} \right\} \\
&=& \int d^4x \sqrt{\gamma_{(0)}}
\left\{ {144 \over \tilde{l}^4}-{12 \over \tilde{l}^4}h_{ij}h^{ij}
+{6\over \tilde{l}^2}h^{ij}\nabla^2 h_{ij}+
{12 \over \tilde{l}^2} \nabla^2 (h_{ij}h^{ij}) \right\}\; .\nonumber
\eea
Expanding the Ricci tensor $R_{ij}$ with respect to $h_{ij}$ 
\be
\label{RicciP}
R_{ij}= R_{(0)ij} + \delta R_{ij} + {1 \over 2} \delta^2 
R_{ij} + {\cal O}\left(h^3\right)
\ee
and  using following expressions,  
\bea
\label{RicciP2}
\delta R_{ij}&=& {1 \over 2}\left( {8 \over \tilde l^2}h_{ij}
- \nabla^2 h_{ij}\right) \nn
\gamma^{ij} \delta ^2 R_{ij} &=&
{4 \over \tilde{l}^2}h_{ij}h^{ij}+\nabla^2 (h_{ij}h^{ij})
-{1\over 2}h^{ij} \nabla ^2 h_{ij}  \; ,
\eea
we obtain
\bea
\label{rrij}
&&\int d^4x \sqrt{\gamma} R_{ij} R_{kl}\gamma^{ki}\gamma^{lj}
=\int d^4x \sqrt{\gamma} \left\{
R_{(0)ij} R_{(0)kl}\gamma_{(0)}^{ki}\gamma_{(0)}^{lj}\right. \nn
&& \qquad + 2\delta R_{ij} R_{(0)kl}\gamma_{(0)}^{ki}\gamma_{(0)}^{lj} 
-2 R_{(0)ij} R_{(0)kl}
\gamma_{(0)}^{kn}h_{nm}\gamma_{(0)}^{mi}\gamma_{(0)}^{lj} \nn
&& \qquad + \delta ^2 R_{ij} R_{(0)kl}\gamma_{(0)}^{ki}
\gamma_{(0)}^{lj} \nn
&& \qquad + 2 R_{(0)ij} R_{(0)kl}
\gamma_{(0)}^{kn}h_{nm}\gamma_{(0)}^{m\tau }h_{\tau \rho}\gamma_{(0)}^{\rho i}
\gamma_{(0)}^{lj} \nn
&& \qquad -4 \delta R_{ij} R_{(0)kl}
\gamma_{(0)}^{km}h_{mn}\gamma_{(0)}^{ni}\gamma_{(0)}^{lj} 
+ \delta R_{ij} \delta R_{kl}\gamma_{(0)}^{ki}\gamma_{(0)}^{lj} \nn
&& \qquad \left. + R_{(0)ij} R_{(0)kl}
\gamma_{(0)}^{kn}h_{nm}\gamma_{(0)}^{mi}
\gamma_{(0)}^{l \tau}h_{\tau \lambda}\gamma_{(0)}^{\lambda j}
\right\} \nn
&& \qquad = \int d^4x \sqrt{\gamma_{(0)}}
\left\{ {36 \over \tilde{l}^4}-{2 \over \tilde{l}^4}h^{ij}h_{ij}
+{1 \over 2 \tilde{l}^2} h^{ij}\nabla ^2 h_{ij} \right.\nn
&& \qquad \left. + {3\over \tilde{l}^2} \nabla^2 (h_{ij}h^{ij})
+{1\over 4}( \nabla ^2 h_{ij})^2 \right\} \; .
\eea
Substituting eqs.(\ref{rr2}) and (\ref{rrij}) into
eq.(\ref{wano}) gives :
\bea
\label{wano2}
&& {\cal W}= \left\{ {2\over 3}b
\left( {1\over 12}+A_{0} \right)+
{1\over 12}\left( b''+{2 \over 3}b' \right) \right\}  
\int d^4x \sqrt{\gamma} \; R^2 \nn
&& \qquad -2b A_{0} \int d^4x \sqrt{\gamma} \; R_{ij}R^{ij} \nn
&& \quad = \left\{ {2\over 3}b
\left( {1\over 12}+A_{0} \right)+
{1\over 12}\left( b''+{2 \over 3}b' \right) \right\}  
\int d^4x \sqrt{\gamma_{(0)}} 
\left\{ {144 \over \tilde{l}^4}-{12 \over \tilde{l}^4}h_{ij}h^{ij}
\right. \nn
&& \quad \left. +{6\over \tilde{l}^2}h^{ij}\nabla^2 h_{ij}+
{12 \over \tilde{l}^2} \nabla^2 (h_{ij}h^{ij}) \right\} 
-2b A_{0} \int d^4x \sqrt{\gamma_{(0)}}
\left\{ {36 \over \tilde{l}^4}-{2 \over \tilde{l}^4}h^{ij}h_{ij} \right. \nn
&& \quad +{1 \over 2 \tilde{l}^2} h^{ij}\nabla ^2 h_{ij} 
\left. + {3\over \tilde{l}^2} \nabla^2 (h_{ij}h^{ij})
+{1\over 4}( \nabla ^2 h_{ij})^2 \right\} \nn
&& \quad = \int d^4x \sqrt{\gamma_{(0)}}\left[
\left\{ b\left( {1\over 18}+ {1\over 6}A_{0} \right)+
{1\over 12}\left( b''+{2 \over 3}b' \right) \right\} 
{144 \over \tilde{l}^4} \right. \nn
&& \quad - \left\{ b\left( {1\over 18}+ {1\over 3}A_{0} \right)+
{1\over 12}\left( b''+{2 \over 3}b' \right) \right\}
{12 \over \tilde{l}^4 }h_{ij}h^{ij} \nn
&& \quad + \left\{ b\left( {1\over 18}+ {1 \over 2}A_{0} \right)+
{1\over 12}\left( b''+{2 \over 3}b' \right) \right\}
{6\over \tilde{l}^2}h^{ij}\nabla^2 h_{ij} \\
&& \quad + \left\{ b \left( {1\over 18}+{1 \over 6} A_{0} \right)+
{1\over 12}\left( b''+{2 \over 3} b' \right) \right\}
{12 \over \tilde{l}^2 } \nabla^2 (h_{ij}h^{ij}) 
\left.-{1\over 2}b A_{0} (\nabla ^2 h_{ij})^2 \right] \nonumber \; .
\eea
Using (\ref{lapl}), one finds the last term 
$-{1\over 2}b A_{0} (\nabla ^2 h_{ij})^2$ gives the leading 
contribution at short distances (that is, large $p$) 
as $(\nabla ^2 h_{ij})^2 \sim p^4 h_{ij}h^{ij}$. Then we find 
that,
by including the effective action ${\cal W}$ in (\ref{Cr3}), 
the term proportional to $p^4$ appears in $T_{p}(y_0)$  
(\ref{ind0}) or $T^{(4)}_{p}(y_0)$  (\ref{Ein1}). Then the 
short distance fluctuation of $h_{ij}$ is highly suppressed 
and the graviton correlator behaves as
\be
\label{Cr4}
\left< h_{ij}(x)h_{i'j'}(x') \right> 
\sim \int {d^4p \over p^4}\e^{ip\cdot\left(x-x'\right)}
\sim \ln\left|x-x'\right|\ .
\ee
In the pure 4d theory, the total action is given by a sum of 
the action  (\ref{actionEin}) and ${\cal W}$ (\ref{wano2})
\be
\label{4dtotal}
S^{(4)}_{\rm total}=S^{(4)}_{\rm EH}  + {\cal W}\ .
\ee
On the other hand, the total action of the 4d gravity induced from 
5d is the sum of the action (\ref{4dcol}) and ${\cal W}$:
\be
\label{totall}
S_{\rm total}=S_{\rm correlator} + {\cal W}\ .
\ee
In both cases, ${\cal W}$ gives the dominant contribution  
$p^4$ at short distances (large $p$) and the pure graviton 
contribution  $p^2$ or $p$ is always 
suppressed by adding ${\cal W}$. Then there does not appear 
the difference between 
the short distance behaviours of the graviton correlator from the 
pure 4 dimensional gravity (\ref{actionEin}) and that of 
the gravity induced from 5 dimensional gravity. 

In \cite{HHR}, the graviton correlator has been found with the help of 
 the higher order surface counterterms, which include the 
square of the curvatures. The counterterms are related with 
the conformal anomaly of the matter  on the brane (in frames of 
 AdS/CFT correspondence). Then, when the radius of 
the 4d de Sitter brane is large, it has been found 
the following graviton correlator in case that the matter 
is ${\cal N}=4$ $SU(N)$ or $U(N)$ super Yang-Mills: 
\bea
\label{HHRcr1}
\left< h_{ij}(x)h_{i'j'}(x') \right> &=& \sum_{p=2}^{\infty}
W_{iji'j'}^{(p)} (x,x') {1\over 2 T^{\rm HHR}_{p}(y_0)}\; , \nn
T^{\rm HHR}_{p}(y_0)&=&{1 \over 8N^2 G_4^2}
\left(p^2 + 3p + 6 + \Psi(p)\right)\ .
\eea
Here $G_4$ is the 4 dimensional Newton constant and $\Psi(p)$ 
is given by
\bea
\label{HHRcr2}
\Psi(p)&=&p(p+1)(p+2)(p+3)\left[\psi\left({p+5 \over 2}\right)
+ \psi\left({p+4 \over 2}\right) - \psi(2) - \psi(1)\right] \nn
&& + p^4 + 2p^3 - 5p^2 - 10p -6 
\eea
with $\psi(z)\equiv {\Gamma'(z) \over \Gamma(z)}$. Then 
$T^{\rm HHR}_p(y_0)$ behaves as $p^4 \ln p$ for large $p$ 
\cite{HHR2}. Neglecting  $\ln p$, the behavior 
seems to be consistent with the results given in this section. 
The formulation is , however, valid for the case that the 
matter on the brane is ${\cal N}=4$ $SU(N)$ or $U(N)$ Yang-Mills 
theory since the higher order counterterms are used in \cite{HHR,HHR2} 
in order to evaluate the quantum contribution from CFT 
 on the brane. The higher order counterterms can be determined 
{\it uniquely} from the geometry of the bulk AdS, which does not 
distinguish what kind of matter exists on the brane. Then the 
corresponding  brane CFT should be always 
${\cal N}=4$ $SU(N)$ or $U(N)$ Yang-Mills theory. Since 
we used the conformal anomaly induced action in order to 
evaluate the contribution from the quantum effects, the 
formulation in this paper can be applied for any 
(conformal) matter theory. 

\section{Discussion}

In summary, we calculated the graviton correlator in
the brane-world model with fixed brane tension where bulk represents
dS space and the brane is (inflationary) de Sitter Universe.
It is shown that like in the standard AdS/CFT related brane-worlds 
the short distance behaviour of the graviton correlator differs 
from that in General Relativity. However, account of quantum effects of
CFT living on the brane drastically changes the short distance behaviour 
of correlator.

In refs.\cite{HHR,NOZ} the quantum induced brane, which is the 4 dimensional 
sphere  in 5-dimensional anti-de Sitter space, has 
been considered. In order for such a brane to exist one 
needs the conformal anomaly induced action (\ref{Cr3}). As a result 
the obtained correlator behaves as (\ref{Cr4}) unlike to the case of 
pure graviton correlator (\ref{Cr2}).     

The short distance behavior (\ref{Cr1}) of the induced 
gravity is different from that of the pure Einstein gravity 
(\ref{Cr2}) if we do not include the quantum effects of the 
matter on the brane. Eq.(\ref{Cr4}) tells, however, 
the quantum effects become dominant at short distances and 
smear the difference. Furthermore the quantum effects 
strongly suppress the fluctuations of the graviton at short 
distances. In other words, the quantum effects make the spacetime 
very rigid on small scales. This is universal feature of quantum CFT 
whatever the gravitational part is. Our discussion gives further support
for consideration of dS brane-worlds on equal footing with AdS brane-worlds.

\section*{Acknowledgments}

The work by S.N. is supported in part by the Ministry of Education, 
Science, Sports and Culture of Japan under the grant n. 13135208.
The work by S.O. is supported in part by the Japan Society
for the Promotion of Science under the Postdoctoral Research Program.

\end{document}